\def \abc#1#2#3#4 {\reference#1, {\sl#2}, {\bf#3}, #4}
\def \blank {\lower 5pt\hbox to 0.75in{\hrulefill}}

\def \cm{~\rm{cm}}
\def \s{~\rm{s}}
\def \km{~\rm{km}}

\def \K{~\rm{K}}

\def \AU{~\rm{AU}}

\def \erg{~\rm{erg}}

\def \yrs{~\rm{yrs}}
\def \yr{~\rm{yr}}
\def \day{~\rm{day}}

\documentclass[12pt,preprint]{aastex}

\begin{document}
\small

\setcounter{page}{1}

\begin{center}
\bf
X-RAY EMISSION FROM CENTRAL BINARY SYSTEMS \\
OF PLANETARY NEBULAE 
\end{center}
\begin{center}
Noam Soker\\
Department of Astronomy, University of Virginia \\
and \\
Department of Physics, University of Haifa at Oranim\\
Oranim, Tivon 36006, ISRAEL; soker@physics.technion.ac.il\\
\bigskip
and \\
\bigskip
Joel H. Kastner \\
Chester F. Carlson Center for Imaging Science \\
Rochester Institute of Technology, 54 Lomb Memorial Dr. \\
Rochester, NY 14623;  jhkpci@cis.rit.edu 
\end{center}
\begin{center}
\bf ABSTRACT
\end{center}
  We study the conditions under which a main sequence binary companion to
the central ionizing star of a planetary nebula (PN) might become 
magnetically  active and thereby display strong X-ray luminosity, 
$L_x \gtrsim 5 \times 10^{29} \erg \s^{-1}$.
 Since most PNe are older than few billion years, any main sequence 
companion will rotate too slowly to have magnetic activity and hence
bright X-ray emission, unless it is spun-up. We demonstrate that
if the orbital separation during the AGB phase of the PN progenitor 
is $a \lesssim 30-60 \AU$, main sequence companions in the spectral type
range F7 to M4 (mass range $0.3 M_\odot \lesssim M_2  \lesssim 1.3 M_\odot$)
will accrete enough angular momentum from the AGB wind to rotate 
rapidly, become magnetically active, and exhibit 
X-ray luminosities $L_x \gtrsim 5 \times 10^{29} \erg \s^{-1}$. 
 Lower mass M stars and brown dwarfs can also become magnetically active,
but they should have small orbital separations and hence are less likely
to survive the AGB phase of the progenitor. 
 For orbital separation of $a \lesssim 0.3 \AU$, i.e., for a
binary systems that went through a common envelope phase, 
the fast wind from the
central WD star will interact with (and potentially disrupt)
the companion's corona on the 
side facing the central star, while for $a \lesssim 6 R_\odot$, i.e.,
an orbital period of $P_{\rm orb} \lesssim 30~$hours, the WD's fast wind will 
compress a dense small region near the surface of the companion. This region
may thermally emit X-rays with nonnegligible luminosity.
We estimate that $20-30 \%$ of elliptical PNe and $30-50\%$ of bipolar PNe
are likely to have magnetically active companions
which will reveal themselves in X-ray observations.  
Re-analysis of Chandra X-ray Observatory spectroscopy of the
compact central source of NGC 7293 indicates that the
emitting region of this object possesses abundance anomalies
similar to those of coronally active main-sequence
stars. High-resolution X-ray spectroscopy of this and other
compact sources in PNe are necessary to confirm a
coronal origin for the X-ray emission.


{\it Subject headings:} planetary nebulae$-$stars: mass loss$-$stars:
magnetic field$-$X-rays: stars

\section{INTRODUCTION}

Planetary nebulae (PNe) -- the ejected, ionized envelopes of expired red
giants -- represent very late stages in the deaths of
intermediate-mass (1-8 $M_\odot$) stars.  Einstein and ROSAT observations
revealed that many PNe are X-ray sources (see Kastner {\it et al.}
2000 and references therein). None of these nebulae was extended
and/or bright enough in X-rays to be resolved spatially by the
Einstein or ROSAT X-ray telescopes. However, a small fraction of the
PN X-ray sources detected by ROSAT appeared too hard ($T_x
> 10^6$ K) for the emission to be ascribed to a recently revealed,
central white dwarf (e.g., Kreysing {\it et al.} 1992, 
Guerrero, Chu, \& Gruendl 2000). 
Prior to the launch of the Chandra X-ray Observatory (CXO),
this relatively high-temperature emission served as the best evidence
for the presence of extended, shock-heated X-ray emission from PNe.

Recent observations by the {\it Chandra} X-Ray Observatory 
(Guerrero {\it et al.} 2001, hereafter GCGWK; Chu {\it et al.} 2001; 
Kastner {\it et al.} 2000; Kastner, Vrtilek \& Soker 2001) have 
established, however, that PNe can in principle 
harbor two types of (relatively) bright, hard X-ray emitting regions: 
diffuse thermal emission from a large cavity inside 
the main nebular shell and a point-like region near the central star.
 In the present paper we consider the nature of these latter, point-like 
X-ray emission sources.

Chandra imaging of the PNe NGC 6543 (the Cat's Eye nebula) and 
NGC 7293 (the Helix nebula) reveal that each possesses a hard
point source, with characteristic temperatures of $\sim 2 \times 10^6 \K$ 
and $ \sim 7 \times 10^6 \K$, and luminosities of 
$\sim 10^{30} \erg \s^{-1}$ and  $\sim 3 \times10^{29} \erg \s^{-1}$, 
respectively (GCGWK).
 In addition the X-ray emission from the central source in NGC 7293
appears time variable (GCGWK). 
 Point sources were not obviously detected in BD +30 3639 or NGC 7027, 
the only other PNs 
thus far imaged by Chandra (the compact source near the central star
of NGC 7027 is most likely a knot of extended emission).
 GCGWK qualitatively discuss four mechanisms for the formation
of the central point-like hard X-ray emission in NGC 6543 and NGC 7293:
\begin{enumerate}
\item Interaction of a fast ($\sim 1000 \km \s^{-1}$) wind from the central 
star with the nebular gas. 
(Although the central ionizing star is evolving to become a WD,
we term it here a WD). 
 GCGWK rule out this possibility since presently
NGC 7293 has no fast wind, and the interaction between the fast wind
and the nebulae gas will form an extended source, as indeed is observed
in NGC 6543 (Chu {\it et al.} 2001), BD+30°3639 (Kastner {\it et al.} 2000),
and NGC 7027 (Kastner {\it et al.} 2001). 
\item
Shocks in the fast stellar wind, similar to
a mechanism suggested to explain X-ray emission from massive
O and B stars.
 GCGWK argue that such a process may occur in NGC 6543.
\item Accretion of material from a close binary companion by the ionizing WD
star. GCGWK rule this out as it is inconsistent with both the X-ray temperature
and the lack of optical time variability. 
\item Coronal (magnetic) activity of a main sequence binary companion.
A dwarf (main sequence) M companion will not be detected due to its low
bolometric luminosity, but it may be magnetically active enough to
emit the observed X-ray luminosity. 
 The time variability of NGC 7293 (GCGWK) and of NGC 6543 in H$\alpha$
(Gruendl {\it et al.} 2001) support the hypothesis of a dMe (dwarf M 
star with strong emission) companion with magnetic activity.
\end{enumerate}
 
 In the present paper we conduct a theoretical study of two
mechanisms by which a main sequence companion to the WD ionizing 
source of the PN can lead to X-ray emission: 
magnetic activity of a  companion (process 4 studied by GCGWK), and 
the collision of the fast wind from the WD with the corona of
a main sequence companion.
 The former magnetic activity mechanism is not to be confused with  
the X-ray point source emission mechanism from the central WD
proposed by Blackman, Frank, \& Welch (2001). 
 We explore the implications of binary interaction since,
as we discuss in \S 2, it is exceedingly likely that a large fraction of
PNe possess close binary central stars.
In $\S2$ we study both the companion magnetic activity and WD-companion
interaction processes. The implications for observations
are discussed in $\S3$, while a short summary is presented in $\S 4$.
 
\section{X-RAY EMISSION FROM BINARY CENTRAL SYSTEMS}
\subsection{Magnetic Activity on a Late Type Companion}

\subsubsection{Requirements for luminous X-ray emission}

 The magnetic activity (e.g., Saar \& Brandenburg 1999) and the X-ray 
luminosity (e.g., Neuh\"auser {\it et al.} 1995) 
of main sequence stars type F7 to M decreases with increasing age or orbital 
rotational period.
 Stars earlier than F7 have very weak magnetic activity (e.g.,
Saar \& Brandenburg 1999). 
 As described later, most surviving active companions to PN central stars
have masses in the range
$0.3 M_\odot \lesssim M_2 \lesssim 1.3 M_\odot$, spectral type  
K4$-$F7. 
 A steep decrease in activity seems to occur at an age of $\sim 10^9 \yr$,
when the X-ray luminosity falls below $10^{29} \erg \s^{-1}$, much below
the luminosities found by GCGWK in the central sources of NGC 6543 and
NGC 7293. 
 To be active, either the companion to the central ionizing star (the WD)
is a very young star, age$~\lesssim 3 \times 10^{8} \yr$, or else
it has been spun-up by accretion and/or tidal interaction. 
 A young companion means fast evolution of the PN progenitor, which
implies a massive progenitor, i.e., $M_i \gtrsim 3 M_\odot$, hence probably
a bipolar PN. 
 Not many PNe belong to this group; most elliptical PNe are formed from 
stars having initial mass of $1 M_\odot  \lesssim M_i  \lesssim 2 M_\odot$, 
hence any main sequence companion of type G to M will be a slowly 
rotating star with low magnetic activity, i.e. 
$L_x \lesssim 3 \times 10^{28} \erg \s^{-1}$. 
 We therefore study the possibility
that the companion is spun-up by accretion. 
 
 We first estimate the rotation period (or velocity) needed to achieve 
high activity, on the level of $L_x \gtrsim 5 \times 10^{29} \erg \s^{-1}$.
We crudely fit the data on the Pleiades, Hyades and the sun given by 
Neuh\"auser {\it et al.} (1995) and Stelzer \& Neuh\"auser (2001)
by 
\begin{equation}
L_x \simeq 10^{29} [3.40-2.37 \log(P/\day)] \erg \s^{-1} \qquad  P < 26 \day,
\end{equation}
where $P=2 \pi / \Omega$ is the rotational period and $\Omega$ the angular
velocity.  
 For long orbital periods we derive a relation based on optical observations
of active stars. 
 From Brandenburg, Saar \& Turpin (1998) we take the relation between
the Ca H and K flux divided by the bolometric luminosity, 
$\left< R^\prime _{\rm HK} \right>$, and the magnetic field, i.e., 
$\left< R^\prime _{\rm HK}\right> \propto (\left< B \right>/B_{\rm eq})^{0.47}$, 
where $\left< B \right>$ is the average magnetic field and $B_{\rm eq}$ is 
the equipartition (with gas pressure) magnetic field value.
 From Saar \& Brandenburg (1999) we take the relation between 
$\left< R^\prime_{\rm HK} \right>$ and the Rossby number 
Ro$\equiv (2 \Omega \tau_c)^{-1}$, that is,
$\left< R^\prime_{\rm HK} \right> \propto Ro^{-1}$, where 
$\tau_c$ is the convective overturn time, and the Rossby number is 
basically the ratio between the rotation period and the convective 
overturn time. 
 Neglecting differences in $B_{\rm eq}$ and $\tau_c$ between different 
types of stars (F7 to M4) we find that $\left< B \right> \propto {\rm Ro}^{-2}$.
Collier, Cameron \& Li (1994), who study magnetic braking, 
argue for $\left< B \right> \propto {\rm Ro}^{-1}$.
 For the purposes of the present study we use the results of 
Saar \& Brandenburg (1999), which are based on magnetic activity. 
 Assuming that the X-ray luminosity is proportional to the magnetic 
energy density $L_x = \left< B \right> ^2$ we obtain the following relation
\begin{equation}
L_x \simeq  2.3 \times 10^{29} \left( \frac{P}{10 \day} \right)^{-4} 
\erg \s^{-1} \qquad  P \gtrsim 7 \day,
\end{equation}
where the scaling was chosen to fit the sun, $P=26 \day$ and 
$L_x= 5 \times 10^{27} \erg \s^{-1}$. 
  Equation (1) and (2) become equal for $P = 13.3 \day$, while 
the derivative $d L_x/d P$ of equations (1) and (2) becomes equal for 
$P=14 \day$, hence we take equation (1) for $P<13 \day$ and equation (2)
for $P \geq 13 \day$. 
 The X-ray luminosity from ROSAT observations of K stars (Wheatley 1998) 
 and of the open cluster IC 4665 (Giampapa, Prosser \& Fleming 1998) 
can be fitted with the same form as equation (1), but with a higher maximum 
activity and a steeper rise with angular velocity
\begin{equation}
L_x \simeq 10^{30} [1- \log(P/\day)] \erg \s^{-1} ,,
\end{equation}
for K stars with periods in the range $0.4 <P < 9.8~$day.
 
 The relevant conclusion from equations (1)-(3) and the papers cited 
in relation to these equations is that, to reach an X-ray luminosity of
$L_x > 5 \times 10^{29} \erg \s^{-1}$, a main sequence companion of 
spectral type F7 to M4 has to be spun up to $P \lesssim 3 \day$.
 For a solar type star the equatorial rotation velocity should be 
$v_{\rm rot} \gtrsim 20 \km \s^{-1}$, or about an order of magnitude 
faster rotation than the present sun has.  
 We take this as the condition on the rotational velocity and period
to induce magnetic activity in companions of PN central stars.

\subsubsection{Accretion by the companion star}

 The mass accretion rate is taken to follow the Bondi-Hoyle relation
$\dot M_a = \pi R_a^2 v_r \rho$, where $R_a=2 G M_2/v_r^2$ is the accretion
radius, and $v_r$ is the relative velocity between the wind and the companion.
 The density is that of the primary wind $\rho = \dot M_1 /(4 \pi a^2 v_{w1})$,
where $\dot M_1$ is the mass loss rate (taken to be positive),
$a$ is the orbital separation, and $v_{w1}$ the primary wind velocity. 
 For accretion from a wind, the net specific angular momentum of the
material entering the Bondi-Hoyle accretion cylinder with radius $R_a$,
i.e., the material having impact parameter $b<R_a$, is 
$j_{BH} = 0.5 (2 \pi / P_o) R_a^2$ (Wang 1981), 
where $P_o$ is the orbital period.
  Livio {\it et al.} (1986; see also Ruffert 1999) find that the actual
accreted specific angular momentum for high Mach number flows is
$j_a = \eta j_{BH}$, where $\eta \sim 0.1$ and $\eta \sim 0.3$
for isothermal and adiabatic flows, respectively.
 The relative velocity is $v_r^2 \simeq v_{w1}^2 + v_o^2$, where $v_{w1}$
is the (slow) wind velocity at the location of the accreting star, and $v_o$
is the relative orbital velocities of the two stars.
 We find below that the orbital separation above which no spin-up occurs
is $> 10 \AU$; at this separation, 
the orbital velocity is lower than the wind speed.
We therefore take $v_r \simeq v_{w1} \simeq 15 \km \s^{-1}$ in what follows. 
 If an accretion disk is formed, then the specific angular momentum of the
accreted mass is that at the inner boundary of the disk, i.e., 
the specific angular momentum of a particle in a Keplerian orbit at
the equator of the accreting star of radius $R_2$ and mass $M_2$, 
$j_2=(G M_2 R_2)^{1/2}$. 
 An accretion disk will be formed if $j_a > j_2$.
  Substituting typical values for main sequence accretor and the mass-losing
terminal AGB star, e.g., a mass of $0.6 M_\odot$ at the end of the AGB,
we find the following condition for the formation of a disk 
\begin{eqnarray}
1< \frac {j_a}{j_2} \simeq 1.3 
\left( \frac {\eta}{0.2} \right)
\left( \frac {M_1+M_2}{1.6 M_\odot} \right)^{1/2}
\left( \frac {M_2}{1 M_\odot} \right)^{3/2}
\left( \frac {R_2}{1 R_\odot} \right)^{-1/2}
\left( \frac {a}{20 \AU} \right)^{-3/2}
\left( \frac {v_r}{15 \km \s^{-1}} \right)^{-4} ,
\end{eqnarray}
where the expression is for a circular orbit with semi-major axis $a$.
 The mass accretion rate decreases strongly with increasing wind speed,
hence most of the mass will be accreted during the AGB phase, when 
the wind speed is low. 
 If the AGB star loses a mass of $\Delta M_{\rm AGB}$, the total accreted 
mass by the companion is 
\begin{eqnarray}
\Delta M_{\rm acc} \simeq  0.04 \Delta M_{\rm AGB} 
\left( \frac {M_2}{1 M_\odot} \right)^2
\left( \frac {a}{20 \AU} \right)^{-2}
\left( \frac {v_r}{15 \km \s^{-1}} \right)^{-4}. 
\end{eqnarray}

The moment of inertia of a companion F7$-$M4 spectral type star is 
$I \simeq 0.1 M_2 R_2^2$, and the rotation velocity is 
$v_{\rm rot}=J_a R_2/I$, where $J_a$ is the accreted angular momentum.
Here we assume solid body rotation, which is
a reasonable assumption for these stars,
which have convective envelopes.
 We first consider accretion from an accretion disk, whose formation is
described by
equation (4) for systems having an orbital separation of $a \lesssim 20 \AU$,
or even $a \lesssim 60 \AU$ for a slow wind, $v_{w1} \simeq 10 \km \s^{-1}$,
at the final stages of the AGB (Soker 2001).  
 Note that the relevant orbital separation is that characterizing
the accretion
phase, and not the initial separation; for systems with $a>20 \AU$ the 
final separation will be larger than the initial one due to mass loss. 
 We find for accretion from a disk
\begin{eqnarray}
v_{\rm rot} \simeq 90 
\left( \frac {\Delta M_{\rm AGB}}{0.5 M_\odot} \right) 
\left( \frac {M_2}{1 M_\odot} \right)^{3/2}
\left( \frac {R_2}{1 R_\odot} \right)^{-1/2}
\left( \frac {a}{20 \AU} \right)^{-2}
\left( \frac {v_r}{15 \km \s^{-1}} \right)^{-4} \km \s^{-1}. 
\end{eqnarray}
 In case a disk is not formed, and the material is accreted directly 
onto the star, the accreted angular momentum is 
$J_a = \Delta M_{\rm acc} j_a$.
 Using the expressions for $j_a$ and $\Delta M_{\rm acc}$ given above 
we find for accretion directly onto the star 
\begin{eqnarray}
v_{\rm rot} \simeq 110  
\left( \frac {\Delta M_{\rm AGB}}{0.5 M_\odot} \right) 
\left( \frac {M_2}{1 M_\odot} \right)^{3}
\left( \frac {R_2}{1 R_\odot} \right)^{-1}
\left( \frac {M_1+M_2}{1.6 M_\odot} \right)^{1/2}
\left( \frac {a}{20 \AU} \right)^{-7/2}
\left( \frac {v_r}{15 \km \s^{-1}} \right)^{-8} 
\left( \frac {\eta}{0.2} \right)        \km \s^{-1}. 
\end{eqnarray}

 Both equations (6) and (7) show that F7$-$M4 spectral type main sequence 
stars in the mass range $0.3 \lesssim M_2 \lesssim 1.3
M_\odot$ that are accreting
from an AGB wind will be spun-up sufficiently
to have strong magnetic activity if
the orbital separation is $a \lesssim 30 \AU$ (with smaller
orbital separations
for lower mass stars) for $v_{w1} \simeq 15 \km
\s^{-1}$. The range of separations may be larger
($a \lesssim 65 \AU$) if, during the final intensive wind phase of the AGB,
the wind velocity is lower, i.e., $v_{w1} \simeq 10 \km
\s^{-1}$ (the total mass  
lost would then be only $\Delta M_{\rm AGB} \sim 0.2 M_\odot$; Soker 2001). 
 Such companions are likely to substantially shape the PN,
such that it becomes an extreme elliptical or even bipolar
nebula (Soker 2001). 
 
 For much smaller orbital separations, $a \lesssim 2 \AU$, the companion
accretes via Roche lobe overflow and/or enters the primary's envelope.
In both cases it will be spun-up substantially.
  Hence, we predict that all close binary systems in PNe should show
magnetic activity if the companions are within the
approximate spectral type range F7 through M, and even brown dwarfs
if survive the evolution.  
 Finally, note  that we do not consider the variation of accretion rate
along eccentric orbits, but rather consider only the semi-major axis $a$; the
uncertainties in other parameters (e.g., the exact
dependence of X-ray luminosity on rotation rate) preclude such a treatment. 

 Low mass main sequence and brown dwarf companions will also
be spun-up via mass accretion during the central star's AGB
phase. However, they need to be at orbital separations of $a \lesssim 10 \AU$
in order to accrete enough angular momentum to become
magnetically active and thereby display bright X-ray emission.
 Many of these objects would not survive to the PN phase
since they would enter the 
AGB envelope by tidal interaction, and then spiral all the
way in and collide with the AGB core. 
 This is the reason for taking the lower companion mass to be 
$\sim 0.3 M_\odot$, although some lower mass companions may survive if 
they are not too close to the envelope, but still close enough to 
accrete enough angular momentum.
 We estimate the initial orbital range for $\sim 0.1-0.3 M_\odot$ 
companions for both surviving and being spun-up to strong magnetic activity 
to be  $5 \lesssim a_i \lesssim 10 \AU$, or up to $\sim 20
\AU$ if the primary  
terminates the AGB with a relatively slow wind ($v_{w1}
\simeq 10 \km \s^{-1}$). 

\subsection{Colliding White Dwarf and Companion Winds}

 In the sun the X-ray luminosity $L_{x , \odot}$ is of the same order 
of magnitude as the rate of kinetic energy carried by the solar wind
$\dot E_k$, $L_{x, \odot} \simeq \dot E_k \simeq 3 \times 10^{27} \erg \s^{-1}$. 
 Collier Cameron \& Li (1994) argue that the mass loss rate from main
sequence stars scales either linearly or as square of the surface 
magnetic field. 
 Based on the behavior of the Sun, we take 
$\dot E_k = L_x$. 
By analogy, the mass loss rate of a main-sequence companion
to a PN central star (defined positively) is then
\begin{eqnarray}
\dot M_2 \simeq 10^{-11} 
\left( \frac {L_x}{10^{30} \erg \s^{-1}} \right)
M_\odot \yr^{-1}.
\end{eqnarray}
 During the early PN phase the central ionizing source ---
the emerging WD ---
blows a fast wind, which will collide with the main sequence companion's wind.
 The stagnation point is where the ram pressures of the two winds are equal, 
i.e., $\rho_f v_f^2 = \rho_2 v_2^2$, where subscripts $f$ and $2$ 
stand for the fast wind from the primary and the wind from the secondary
main sequence star, respectively. 
 Substituting for the density and scaling with typical values
we obtain
\begin{eqnarray}
d_2 \simeq 0.016  a 
\left( \frac {L_x}{10^{30} \erg \s^{-1}} \right)^{1/2}
\left( \frac {\dot M_f v_f}{2 \times 10^{-5} M_\odot \yr ^{-1} \km \s^{-1}}
\right)^{-1/2}
\left(\frac{v_2}{500\km\s^{-1}}\right)^{1/2}, 
\end{eqnarray}
where $d_2$ is the distance of the stagnation point from the companion's 
center.
 The electron density $n_{2e}$ of the companion's shocked wind at the 
stagnation point is four times the free wind density,   
\begin{eqnarray}
n_{2e} \simeq 9 \times 10^4   
\left( \frac {a}{20 \AU} \right)^{-2} 
\left( \frac {\dot M_f v_f}{2 \times 10^{-5} M_\odot \yr ^{-1} \km \s^{-1}}
\right) 
\left(\frac{v_2}{500 \km \s^{-1}}\right)^{-2} \cm^{-3}.
\end{eqnarray}
The shocked secondary wind gas temperature is $T \simeq 3.5 \times 10^6 \K$, 
and its cooling time is 
\begin{eqnarray}
\tau_c \simeq 20 \left( \frac {a}{20 \AU} \right)^{2} \yr.
\end{eqnarray}
 The flow time of the shocked gas out from the interaction region is
$\tau_f \sim d_2/v_2$ for $d_2 \gtrsim R_2$, 
and $\tau_f \sim R_2/v_2$ for $d_2 \lesssim R_2$.
 For $R_2 = R_\odot$, the mass loss rate from the secondary as 
determined from equation (8),  and a WD wind of 
$\dot M_f v_f = 2 \times 10^{-5} M_\odot \yr ^{-1} \km \s^{-1}$,
we find $d_2 = R_2$ for $a \simeq  0.3 \AU$.  
 For $d_2 \gtrsim R_2 \simeq R_\odot$ the flow time is much shorter than
the cooling time, hence the shocked secondary wind will radiate only
a tiny fraction of its thermal energy before flowing away from the 
wind interaction region and expanding, hence increasing farther its 
cooling time.
 In that case the thermal X-ray emission from the shocked secondary
wind is negligible.

 We next examine the situation when the two stars are very close,
and the WD wind compresses the corona of the main sequence companion.
 The density is determined by the ram pressure of the WD fast wind, hence
it is still given by equation (10) and the cooling time is given
by equation (11).
 The flow time is $\tau_f \sim R_2/v_2 \simeq 20~$minutes. 
 The flow time will be longer than the cooling time for  
$a \lesssim 6 R_\odot$, which corresponds to an orbital period of 
$P_{\rm orb} \lesssim 30~$hours.
 The condition for significant X-ray emission from the 
thermalized outflow from the companion is therefore 
\begin{eqnarray}
a \lesssim 6
\left( \frac {\dot M_f v_f}{2 \times 10^{-5} M_\odot \yr ^{-1} \km \s^{-1}}
\right)^{1/2} R_\odot.
\end{eqnarray}
 We note the following properties of this compression process. 
\newline
(1) The WD wind compresses its main sequence companion's
corona on the side facing the WD,
to a very thin ``shell''. 
 This is true already for $a \lesssim 0.3 AU= 65 R_\odot$ 
($P_{\rm orb} \lesssim 0.13 \yr$), as noted above.  
 In this case we expect the magnetic activity (and therefore
flaring) on the side of the companion
facing the WD to be disrupted. It is uncertain as to whether
the companion's X-ray luminosity would decrease or increase
given such conditions. 
\newline
(2) For $a \lesssim 6 R_\odot$ the electron density in the compressed corona
on the side facing the WD reaches values of 
 $n_{2e} \gtrsim 5 \times 10^{10} \cm^{-3}$. In this case
the cooling time is shorter 
than the flow time, and the gas manages to emit most of its thermal energy. 
At these densities, collisional de-excitation starts to be important, 
influencing the X-ray spectra. We stress, however, that the processes in 
close binary stars are much more complicated (e.g., Stepien, Schmitt, \& 
Voges 2001)
and a full treatment is beyond the scope of this paper.
\newline
(3) The volume occupied by the dense gas is very small. 
To achieve an X-ray luminosity of 
$L_x({\rm wind)}= 5 \times 10^{29} \erg \s^{-1}$
on the side facing the WD, the shell thickness should be only 
 $h_{\rm shell} \simeq  0.003 (a/ 6 R_\odot)^{4} R_\odot$,
for the fast wind parameters assumed above. 
 This implies that the coronal structure will be totally disrupted by
the WD winds, and that gasdynamical calculations are required
before our simple estimates of the X-ray emission from the thermalized
gas can be fully trusted.  

\section{OBSERVATIONAL IMPLICATIONS}
\subsection{The Fraction of PNe with Magnetically Active Binary Systems}

 In this subsection we estimate the fraction of PNe that we expect to
possess a central binary system where the companion show magnetic activity.  
 Due to the many uncertainties, e.g., the AGB wind velocity at
the final intensive wind phase (Soker 2001), the exact condition to
set magnetic activity in old stars, and the percentage of PNe formed from
binary systems, our estimates are crude, but still may be meaningful for
future observations. 

 Soker \& Rappaport (2000) estimate that $\sim 10 \%$ of all PNe have 
tidal-strongly interacting binaries in which the secondary
stays outside the AGB envelope  
most of the evolutionary time. Most of these systems form bipolar PNe.
 Most such binaries contain a main sequence companion, with half of 
spectral type F7$-$M4; such binaries therefore constitute
$\sim 4\%$ of all PNe.  
 Another few percent of bipolar PNe are formed by common envelope evolution.
Hence, in total, $\sim 15 \%$ of all PNe are bipolar, and in
$\sim 5-8\%$ of these cases (i.e., $30-50\%$ of all bipolar PNe)
we expect a magnetically active main sequence companion to
be present.
 
 From a population synthesis study, Soker (2001) found that $5 - 20\%$
of all PNe possess binary central stars wherein the
companions accrete from the AGB  
wind and form an accretion disk, although the tidal interaction stays
weak along the entire evolution (the lower value is for a fast AGB wind 
while the higher one is for a slow AGB wind). 
 In half of these systems the companion is a WD, while in
half (i.e., $\sim 7\%$ of all PNe) the companion is a main sequence star. 
 Hence, we estimate that $\sim 5 \%$ of all PNe central
stars possess a companion that is an F7$-$M4 main sequence star 
which has been spun-up to be X-ray active via wind accretion. 
 These are mainly elliptical PNe, in which the
tidal interaction (between the companion and AGB star) is weak. 
 From the population synthesis studies of 
Yungelson, Tutukov, \& Livio, (1993), Han, Podsiadlowski, \& Eggleton (1995),
and Soker \& Rappaport (2000), we estimate that another 
$\sim 20-25 \%$ of all PNe formed via a common envelope interaction and form
elliptical PNe. Most of these have a companion in the F7$-$M4
mass range that we have determined here to be X-ray emission candidates.
However, many of the central binaries in these PNe suffer
mergers, such that the companion no longer exists. 
Bond \& Livio (1990) and Bond (2000) estimate that $\sim 10-15 \%$ of
all PNe have close binary systems in their center. 
Based on these we estimate a contribution of $\sim 10-15 \%$ of all PNe
from these systems to the population of PNe with 
magnetically active main sequence stars.
 Overall, therefore, we estimate that  $\sim 15-20 \%$ of
all PNe ($\sim 20-30 \%$ 
of elliptical PNe) are elliptical and have magnetically active 
main sequence companions.   
  
 From the theoretical study by Yungelson {\it et al.} (1993)
and observations 
of close binary systems in PNe (Bond \& Livio 1990; Bond 2000) we estimate 
that for $\sim 5-10 \%$ of all PNe, most (but not all) of
which are elliptical,  the orbital separation is 
$6 R_\odot \lesssim a \lesssim 65 R_\odot$.
Hence the magnetic activity on the side of the companion
facing the WD is disrupted, if the latter still blows a fast wind.
 However, many of these WDs would be in a phase where the
fast wind has diminished and no longer influences the
companion's magnetic activity.  
In another $\sim 5-10 \% $ of all PNe, most of which are elliptical,
the orbital separation is $a \lesssim 6 R_\odot$, and we expect significant
X-ray emission from the compressed corona, in the subset for
which the central WD still has a fast wind (i.e., 
$\sim3-7 \%$ of all PNe).

\subsection{X-ray Spectra}

If the X-ray emission from compact sources at the cores of PNe is
due to coronal activity, then we expect this emission to
resemble spectroscopically that of stellar coronae. Such
emission is characterized by an optically thin,
line-dominated spectrum at temperatures of $\sim 10^7 \K$;
typical electron densities lie in the
range $\log {n_e} \sim 9-12$, and abundance anomalies are
commonplace (Linsky 2001). The high temperature
alone is sufficient to distinguish such coronal
emission from the emission due to a hot white dwarf at a
PN's core (Guerrero {\it et al.} 2000; GCGWK).

Existing medium-resolution (CCD) X-ray spectra
may provide some clues to the origin of the emission from PN
cores. Here we present a re-analysis of one such dataset,
for the central source of NGC 7293, obtained with the
Chandra Advanced CCD Imaging Spectrometer (ACIS). These data
(Observation ID 631) were first obtained and analyzed by GCGWK.
Subsequent to publication of this
analysis, the Chandra X-ray Center (CXC) released improved
spectral calibration files for ACIS CCD S3 (Edgar 2001), with which the
NGC 7293 data were obtained, prompting our
re-analysis. We used standard data processing tools available as part of
the CXC CIAO
\footnote {http://asc.harvard.edu/ciao/}
package (version 2.1) to reprocess the Level 1 X-ray events
and produce recalibrated Level 2 events. We then
extracted the spectrum of the central
source from the recalibrated Level 2 events, and constructed
corresponding spectral response
files. The source and spectral response file extraction
region was defined as a $2''$ radius
circle centered on the core point source of NGC 7293; a
background spectrum was extracted from an annulus extending
between $2''$ and $6''$ radius, and subtracted from the
source spectrum.

The resulting spectrum of the NGC 7293 point source is
displayed in Figure 1 (1a for solar abundances, and 1b for
non-solar abundances).
Note that, with re-calibration, the
spectrum is revealed to be somewhat harder than indicated in
Fig. 2 of GCGWK, peaking between 0.8 and 1.0 keV
(rather than between 0.7 and 0.85 keV).
This suggests that Ne emission is anomalously bright in
the NGC 7293 spectrum. Hence, we attempted spectral fitting
with the MEKAL plasma model (Mewe, Lemen, \& van den Oord
1986) using both solar abundances and non-solar
abundances, where in the latter case we varied the
abundances of Ne and other
elements (Fe, Mg, O) with prominent emission lines that fall
within the energy range of
bright emission from NGC 7293.

Results of this fitting
exercise are illustrated in Fig.\ 1. In the cases of both
solar and non-solar abundances, the best-fit temperature is
found to be $kT_x \sim0.7$ keV for an assumed column density of
$N_H = 4\times10^{20}$ cm $^{-2}$, the value determined by
GCGWK (we find that the fits are not sensitive to
the value of $N_H$). It can be seen, however, that a
model holding abundances at solar produces a fit slightly
inferior to that of the non-solar abundance model. In the
latter model, we find best-fit abundances for O, Ne, Fe, and
Mg of 0.1, 1.2, 0.3, and 0.3 relative to solar (with formal
uncertainties of $\pm20$\%). Note that neither model is able
to account for the apparent excess emission at $\sim1.1$
keV, which may be due to anomalously bright Fe L emission
(our attempts to increase the Fe abundance to account for
this emission result in a large overestimate of the emission
intensity at $\sim 0.9~$keV).

The abundances determined in the above fitting exercise are
similar to those characteristic of the
so-called ``inverse first ionization potential effect'' that
is commonly observed in stellar coronae (e.g., Drake {\it et al.} 2001).
Hence these results would appear to support a coronal origin
for the X-ray emission from NGC 7293. However, we caution
that the X-ray spectra of extended nebular emission also
display prominent line emission and are characterized by
abundance anomalies (Chu {\it et al.} 2000; Kastner
{\it et al.} 2000, 2001). Thus, the presence
of such anomalies is not sufficient, in and of itself,
to conclude that compact X-ray sources within PNe are
coronal in nature. High-resolution spectroscopy --- such as
that enabled by the Chandra X-ray Observatory
High Energy Transmission Gratings and the X-ray Multiple
Mirror Reflection Gratings Spectrometer --- is needed,
to determine more precise abundances as well as electron densities. The
latter may represent the most conclusive available test of the coronal
model (Kastner {\it et al.} 2002).

\section{SUMMARY}

 GCGWK argue that the unresolved hard X-ray emission from the center
of the PN NGC 7293 comes from a dMe magnetically active companion,
and that a similar source may explain the hard X-ray emission from the
center of the PN NGC 6543.
 Motivated by these findings --- and by the common view that stellar
companions shape many PNe and that many of these companions
will be on the main sequence ---
we study the conditions for X-ray emission from magnetically active,
spun-up main sequence (and even brown dwarf) stellar companions
to central stars of planetary nebulae (PNe).
 Our main results can be summarized as follows.

(1) We estimate that companions of spectral type F7$-$M4 must have
an equatorial rotational velocity of $\gtrsim 20 \km \s^{-1}$
in order to have sufficient magnetic activity to generate
X-ray luminosities on the order of those observed for the point
sources within PNe, i.e., $L_x \gtrsim 5 \times 10^{29} \erg \s^{-1}$.
Since most PNe are older than a few billion years, any main sequence
companion will rotate much more slowly than this; therefore, we conclude
that X-ray emitting, main-sequence companions have been spun up, most
likely
via accretion of mass from the former AGB star that generated the PN.
  We concentrated on companions in this spectral type range (F7$-$M4)
because
companions earlier than F7 will not show magnetic activity, while low
mass stars later than M4 need to be close to the AGB progenitor
in order to accrete enough mass from the wind to be spun-up.
  Such very low mass
companions are likely to enter the envelope of the AGB star (i.e., they
will undergo a common envelope phase) and are therefore
unlikely to survive.

(2) We found that main sequence companion stars will be spun up (by
acquiring
angular momentum from the AGB progenitor's wind) if the orbital
separation during the AGB phase is $a \lesssim 30 \AU$
(orbital period of $P_{\rm orb} \lesssim 130 \yrs$ for a binary
combined mass of $1.6 M_\odot$) and a wind speed of
$\sim 15 \km \s^{-1}$.
For a slower AGB wind of $\sim 10 \km \s^{-1}$, the condition is
$a \lesssim 65 \AU$ (orbital period of
$P_{\rm orb} \lesssim 400 \yrs$).
Late K to M stars, and even brown dwarfs, need to have somewhat smaller
orbital separations to be spun-up and thereby become magnetically active.
 Much closer companions will be spun-up via accretion from a Roche lobe
overflow or by common envelope evolution.
 Binary systems that go through a common envelope phase end
with much closer orbital separations.

(3) If the final orbital separation is $a \lesssim 65 R_\odot$
($P_{\rm orb} \lesssim 0.13 \yr$) then the fast wind from the central
ionizing source (the eventual WD), if still active, is likely to
influence the magnetic activity of the companion on the side
facing the ionizing source.
 It is not clear if this will increase or decrease the X-ray emission.

(4) If the final orbital separation is $a \lesssim 6 R_\odot$
($P_{\rm orb} \lesssim 30~$hours) the fast wind from the central WD star
will compress a very dense region close to the companion side facing the
WD.

(5) We crudely estimated that $20-30 \%$ of elliptical PNe and
$30-50\%$ of bipolar PN are likely to have magnetically active companions
which will reveal themselves in X-ray observations.
 If we take bipolar and elliptical PNe to be $\sim 15 \%$ and
$\sim 75 \%$ of all PNe, respectively, we find that $20-30 \%$
of all PNe are expected to harbor magnetically active, and therefore
X-ray luminous, main sequence companions.

(6) We reanalyzed Chandra X-ray Observatory spectroscopy of the
compact central source of NGC 7293 (previously studied by GCGWK).
This modeling, while inconclusive,
suggests that the emitting region of this object possesses
abundance anomalies similar to those of coronally active main-sequence
stars, with Ne somewhat overabundant and Fe underabundant relative to
solar.
High-resolution X-ray spectroscopy of this and other
compact sources in PNe are necessary to confirm a
coronal origin for the X-ray emission.

\bigskip

{\bf ACKNOWLEDGMENTS:}
This research was supported in part by grants from the
US-Israel Binational Science Foundation,
and by NASA/Chandra X-ray Center (Smithsonian Astrophysical
Observatory) grant GO0--1067X to RIT.


\begin{figure}
\figurenum{1a}
\plotone{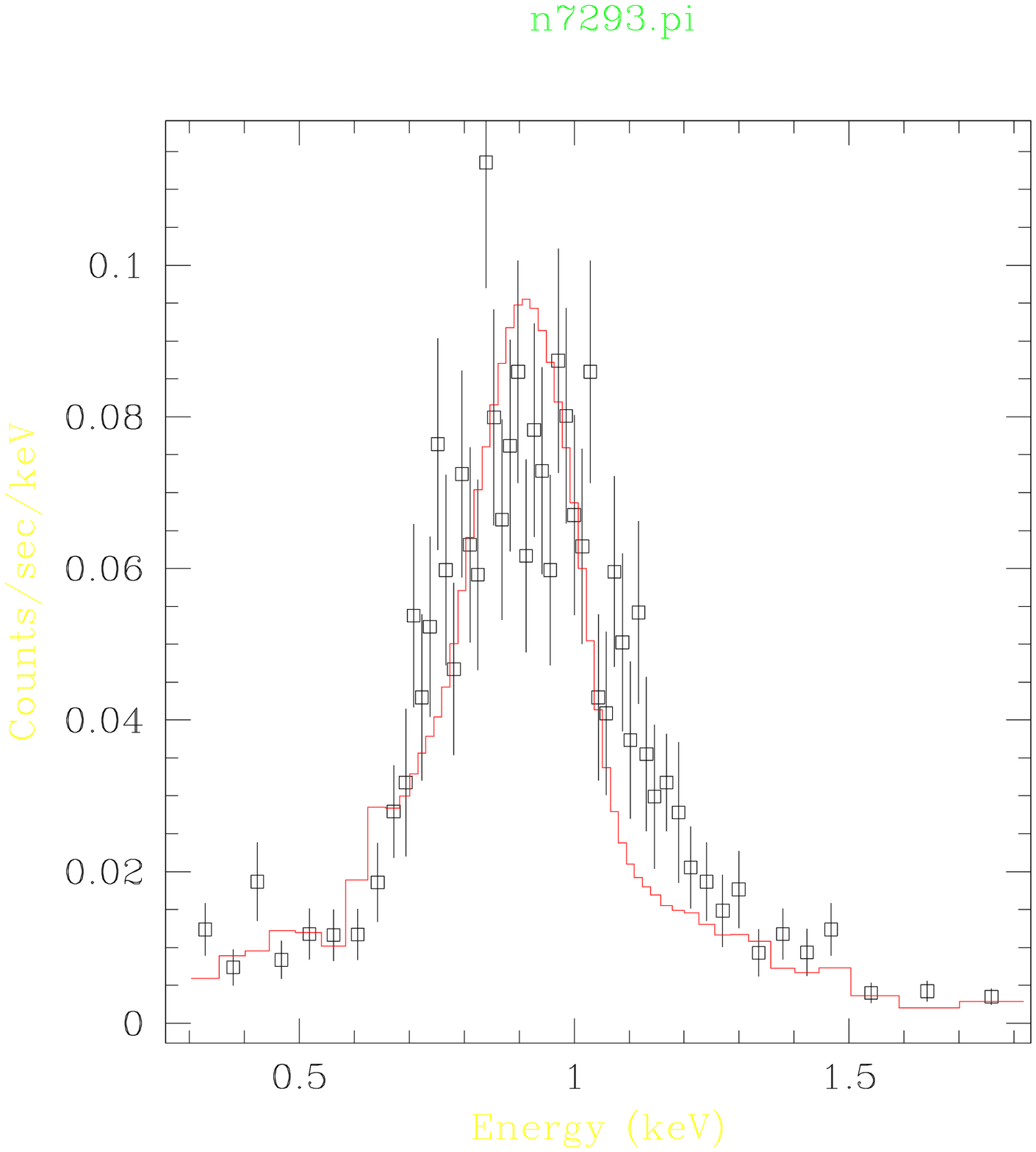}
\caption{
Chandra/ACIS spectrum of the central compact source of NGC
7293, overlaid with the results of fitting MEKAL models with
solar abundances (a), and non-solar abundances (b). In
the latter model, emission from Ne is enhanced, while
emission from O, Mg, and Fe is somewhat suppressed.
}
\end{figure}
\begin{figure}
\figurenum{1b}
\plotone{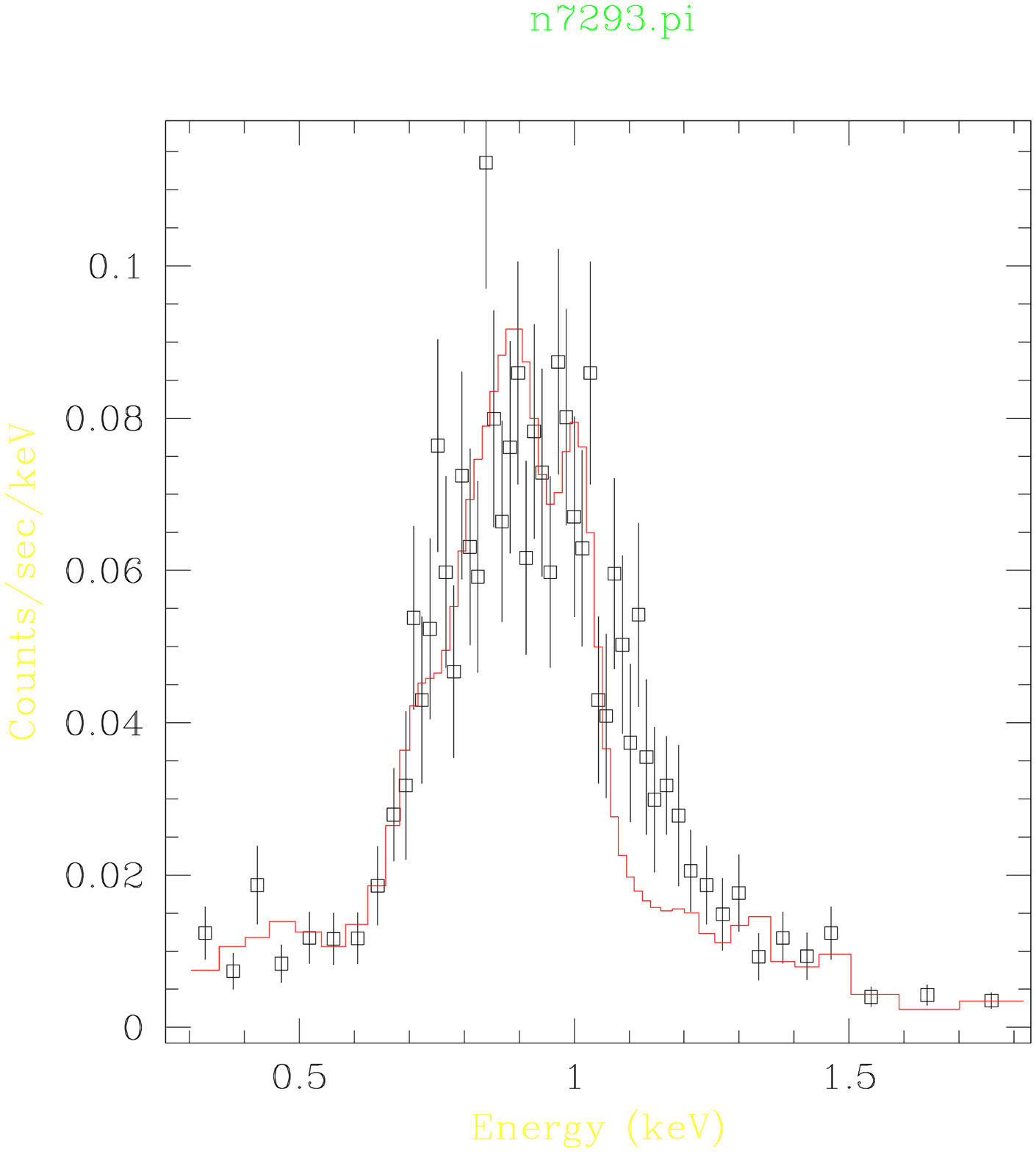}
\caption{
}
\end{figure}

\end{document}